\documentclass[superscriptaddress,twocolumn,pra,10pt,longbibliography]{revtex4-1} 
\usepackage{natbib}
\usepackage{epsfig}
\usepackage{bm}
\usepackage{amsmath}
\usepackage{subfigure}
\usepackage{graphicx}
\usepackage[english]{babel}

\newcommand{\bea}{\begin{eqnarray}}
\newcommand{\eea}{\end{eqnarray}}
\newcommand{\beq}{\begin{equation}}
\newcommand{\eeq}{\end{equation}}
\newcommand{\bes} {\begin{subequations}}
\newcommand{\ees} {\end{subequations}}

\newcommand{\rmi}{{\text i}}

\newcommand{\ignore}[1]{}

\begin{document}

\title{Quantum Annealing for Constrained Optimization}

\author{Itay Hen}
\affiliation{Information Sciences Institute, University of Southern California, Marina del Rey, CA 90292, USA}
\affiliation{Center for Quantum Information Science \& Technology, University of Southern California, Los Angeles, California 90089, USA}
\email{itayhen@isi.edu}

\author{Federico M. Spedalieri}
\affiliation{Information Sciences Institute, University of Southern California, Marina del Rey, CA 90292, USA}
\affiliation{Center for Quantum Information Science \& Technology, University of Southern California, Los Angeles, California 90089, USA}
\affiliation{Department of Electrical Engineering, University of Southern California, Los Angeles, California 90089, USA}


\begin{abstract}
Recent advances in quantum technology have led to the development and
manufacturing of experimental programmable quantum annealers
that promise to solve certain combinatorial optimization problems of practical relevance faster than their classical analogues. The applicability of such devices for many theoretical and real-world optimization problems, which are often constrained, is severely limited by the sparse, rigid layout of the devices' quantum bits. Traditionally, constraints are addressed by the addition of penalty terms to the Hamiltonian of the problem, which in turn requires prohibitively increasing physical resources while also restricting the dynamical range of the interactions. Here, we propose a method for encoding constrained optimization problems on quantum annealers that eliminates the need for penalty terms and thereby reduces the number of required couplers and removes the need for minor embedding, greatly reducing the number of required physical qubits. We argue the advantages of the proposed technique and illustrate its effectiveness. We conclude by discussing the experimental feasibility of the suggested method as well as its potential to appreciably reduce the resource requirements for implementing optimization problems on quantum annealers, and its significance in the field of quantum computing. 
\end{abstract}


\maketitle

\section{Introduction}\label{sec:intro}

Many problems of theoretical and practical relevance consist of searching for the global minima of a cost function defined over a discrete configuration space. These combinatorial optimization problems are not only notoriously hard to solve but are also ubiquitous, appearing in a wide
range of diverse fields such as machine learning, materials design, software verification and logistics, to name a few examples~\cite{papadimitriou2013combinatorial}. It is no surprise then that the design of fast and practical algorithms to solve these has become one of the most important challenges of many areas of science and technology.

One of the more novel approaches to optimization is that of Quantum annealing (QA)~\cite{Finnila1994343,Brooke30041999,kadowaki:98,farhi:01,santoro:02}, a technique that utilizes gradually decreasing quantum fluctuations to traverse the barriers in the energy landscape in search of global minima of complicated cost functions. As an inherently quantum technique, quantum annealers hold the so-far-unfulfilled promise to solve combinatorial optimization problems faster than traditional `classical' 
algorithms~\cite{young:08,young:10,hen:11,hen:12,farhi:12}. 
With recent advances in quantum technology, which have led to the development and
manufacturing of the first commercially available experimental programmable quantum annealing optimizers containing
hundreds of quantum bits~\cite{johnson:11,berkley:13}, interest in quantum annealing has increased dramatically due to the exciting possibility
that real quantum devices could solve classically intractable problems of practical importance. 

One of the main advantages of QA is that it offers a very natural approach to solving discrete optimization problems. 
Within the QA framework (often used interchangeably with the quantum adiabatic algorithm despite certain subtle differences), the solution to an
optimization problem is encoded in the ground state of a problem Hamiltonian
$H_p$. The encoding is normally readily done by expressing the problem at hand as an Ising model, which has a very simple physical interpretation as a system of interacting magnetic dipoles
subjected to local magnetic fields. 

To find the solution, QA prescribes the following course of action. As a first
step, the system is prepared in the ground state of another Hamiltonian
$H_d$, commonly referred to as the driver Hamiltonian.  The driver
Hamiltonian is chosen so that it does not commute with the problem
Hamiltonian and has a ground state that is fairly easy to prepare. 
As a next step,
the Hamiltonian of the system is slowly modified from $H_d$ to
$H_p$, using a (usually) linear interpolation, i.e.,
\begin{equation}\label{eq:hs}
H(s)=s H_p +(1-s) H_d \,,
\end{equation}
where $s(t)$ is a parameter varying smoothly with time 
from $0$ at $t=0$ to $1$ at the end of the algorithm, at
$t=\mathcal{T}$.  If this process is done slowly enough, the
adiabatic theorem of quantum mechanics~\cite{kato:51,messiah:62}
ensures that the system will stay close to the ground state of the
instantaneous Hamiltonian throughout the evolution, so that one finally
obtains a state close to the ground state of $H_p$.  At this point,
measuring the state will give the solution of the original problem with high
probability. 
The running time $\mathcal{T}$ of the algorithm determines the
efficiency, or complexity, of the algorithm and should be large compared to the inverse of the minimum gap~\cite{kato:51,jansen:07,lidarGap}. 

In recent years it has become clear that while QA devices are naturally set up to solve unconstrained optimization problems, they are severely limited when required to solve discrete optimization problems that involve constraints, i.e., when the search space is restricted to a subset of all possible input configurations (normally specified by a set of equations). 

The standard canonical way of imposing these constraints consists of squaring the constraint equations and adding them as penalties to the objective cost function with a penalty factor,  transforming the constrained problem into an unconstrained one~\cite{gaitan:11,gaitan:13,rieffel:14,gaitan:14,lucas:14,rieffel:15,zick:15}. 
This `penalty based' method has several shortcomings. First, the squaring of the constraints
normally results in a pattern of interactions that pairwise couple all the input variables (an `all-to-all' connectivity). 
When mapped to a set of qubits, this translates to the technically unrealistic requirement that the annealer admits programmable interactions between all pairs of qubits~\footnote{
In realistic experimental devices, qubits are typically laid out on a two-dimensional surface, and are forced by engineering limitations to interact only with a limited number of 
neighboring qubits.}. The all-to-all connectivity requirement may be circumvented by `minor embedding' the complete graph onto the hardware graph, however this process is not only very resource demanding, but it is also approximate in nature and is known to reduce the dynamical range of the qubit couplings considerably, while introducing additional constraints~\cite{choi:08,vinci:15}.  Furthermore, due to the required all-to-all connectivity, the energy scale of the added penalty terms imposing the constraints must typically be quadratically larger than the scale of the original problem, which quickly overwhelms the dynamical range of the programmable interactions.

In this work we propose an altogether different approach to solving constrained optimization problems via quantum annealing in which we utilize suitably tailored driver Hamiltonians. 
Within our approach, the addition of penalty terms and all resulting complications become unnecessary. We will show that, when applicable, the annealing itself naturally takes place only in the subspace of configurations that satisfy the constraints and that the amount of resources required for the encoding of a problems reduces dramatically. 

\section{Constrained quantum annealing (CQA)}

We now present the basic principles of our approach.  Let us consider an $n$-qubit problem Hamiltonian encoding a general (classical) discrete optimization problem, 
$H_p(\{\sigma_i^z\})$, where the dependence of  $H_p$ on $\{ \sigma_i^z\}_{i=1}^n$, the set of Pauli $z$-operators acting on the various qubits of the system, could be arbitrary. Let us also assume that the system is
subject to a constraint $C(\{\sigma_i^z\})=c$, where $C$ is also assumed arbitrary and $c$ is a real-valued constant (this approach trivially extends to the case of multiple constraints). 

Traditionally, imposing the constraint would consist of adding the `penalty' term $H_{\text{penalty}}=[C(\{\sigma^z_i\})-c]^2$ to the Hamiltonian, modifying it to $H'_p=H_p + \alpha H_{\text{penalty}}$, where $\alpha$ is a suitably chosen positive constant~\cite{gaitan:11,gaitan:13,gaitan:14,lucas:14,rieffel:15,zick:15}. As already discussed above, the addition of a penalty term is detrimental in several ways. 
First, it requires additional (normally two-body) interactions in the 
problem Hamiltonian, and since in practice actual devices would not be able to accommodate these, costly minor embedding techniques would be in order~\cite{choi:08,vinci:15}. Furthermore, the requirement that ground states of $H_p$ map to those of $H'_p$ introduces an `extra energy scale' to the cost function which in practice translates to increased error levels in the encoding of the couplings.  

Here, we suggest imposing constraints in a somewhat different manner. This is accomplished by first noting that 
since the constraint is `classical,' as an operator it trivially commutes with the problem Hamiltonian, namely, $[H_p,C]=0$.
If one were to find a suitable driver Hamiltonian that also commutes with the constraint, namely $[H_d,C]=0$ (while still not commuting with $H_p$), the constraint would become a \emph{constant of motion}, i.e., would obey $[H,C]=0$ at every instant of time $t$ during the course of the annealing. By further setting up the initial state of the system to be the ground state of the driver Hamiltonian in the relevant sector $\langle C(\{\sigma^z_i\})\rangle_{t=0}=c$,  the dynamics
will naturally take place in the subspace of `feasible' configurations of the optimization problem, automatically obeying the constraint.

Clearly, finding and setting up a suitable driver Hamiltonian for every conceivable constraint may be difficult. However, as it turns out, the most common constraints that appear in the formulation
of classically intractable combinatorial optimization problems have the form of a linear equality (see~\cite{lucas:14} for numerous examples), specifically 
$C(\{\sigma^z_i\})=\sum_{i=1}^n \sigma_i^z = c$. 

Let us now consider now the driver Hamiltonian
 \beq
 \label{eq:driver}
 H_d=-\sum_{i=1}^n \left( \sigma_i^x \sigma_{i+1}^x + \sigma_i^y \sigma_{i+1}^y\right) \,,
 \eeq 
where the label $i=n+1$ is identified with $i=1$. This driver is a special case of the well-known XY-model
which was analytically solved in $1961$ by Lieb, Schultz and Mattis~\cite{lieb:61,pasquale:2008}. This driver has the following attractive properties: i) as can easily be verified, it obeys $[H_d,\sum_{i=1}^n \sigma_i^z ]=0$; ii) furthermore, it is a gapped model, having a non-degenerate ground state, with a gap that scales as $\sim 1/n$~\footnote{
Note that a closing initial gap, albeit somewhat unusual, is not expected to be a bottleneck of the computation, as the minimum gap is expected to close at a much faster rate.}; iii) it requires only basic two-body interactions, and only $n$ of those; iv) finally, the XY model is not only exactly solvable, but may be mapped, via a Jordan-Wigner transformation, to a system of \emph{noninteracting} spinless fermions. As such, it admits a very simple energy landscape, rendering its ground states in the various conserved sectors easily preparable. 

We note that within our constrained quantum annealing (CQA) approach, evolution takes place in the subspace of the full Hilbert space corresponding to those states belonging to the appropriate sector of the conserved quantity. Transitions to other sectors of the conserved quantity are hence suppressed. 
It is therefore appropriate to define the notion of \emph{relevant gap}, which is the gap between the ground state and first excited state \emph{within the relevant sector}. The relevant gap is the one with respect to which the adiabatic running time is to be estimated. Additionally, it is imperative that the choices of driver Hamiltonian and initial state do not `over-constrain' the evolution, restricting it to a smaller subspace than the one prescribed by the constraint. 

\section{Examples}
\label{examples}

We now turn to illustrate the effectiveness of CQA by considering several examples set up to showcase both the generality of the method and its significance in reducing the amount of resources needed to embed optimization problems on quantum annealers.

\subsection{Graph Partitioning}
\label{GP}

The first problem we tackle is that of graph partitioning (GP), where one considers an undirected graph $G=(V,E)$ with an even number $n=|V|$ of vertices and is then asked to find a partition of $V$ into two subsets of equal size $(n/2)$ such that the number of edges connecting the two subsets is minimal.   
GP is known to be an NP-hard problem, where the corresponding decision problem is NP-complete~\cite{karp:72}.   As an Ising model, the GP problem can be written as 
 \bea
H'_p&=&H_p + \alpha H_{\text{penalty}}\\\nonumber
&=&\frac1{2} \sum_{(ij) \in E} \left(1-\sigma_i^z \sigma_j^z\right) + \alpha \left( \sum_{i=1}^n \sigma_i^z \right)^2 \,.
\eea
where the first term assigns positive cost to each edge that connects vertices belonging to different partitions, and the second is a constraint of the form $ \left[C(\{\sigma^z_i\})-c\right]^2$,
 with $C(\{\sigma^z_i\})=\sum_{i=1}^n \sigma_i^z$ and $c=0$, that penalizes unequal partitions. The penalty factor $\alpha$ must obey $\alpha \ge \min(2\Delta, n)/8$ where $\Delta$ is the maximal degree of $G$~\cite{lucas:14}. It is easy to see that 
$H_{\text{penalty}}$ requires a complete (i.e., a fully connected) interaction graph. Furthermore, the energy scale
associated with the penalty term scales with the size of the problem (unless the maximal degree of the graph is bounded).

The above formulation may assume the usual transverse-field driver Hamiltonian $H_d=-\sum_{i=1}^n \sigma_i^x$. However, choosing instead the driver presented in Eq.~(\ref{eq:driver}), the constraint commutes with the total Hamiltonian, Eq.~(\ref{eq:hs}) (here, $H_p$ does not contain any penalties). 
Thus, the instantaneous eigenstates of  $H$ are also eigenstates of $C(\{\sigma^z_i\})$: 
if the initial state is an eigenstate of $C(\{\sigma^z_i\})$ with eigenvalue $c$, so will be the final state; i.e., the constraint in this case is enforced by conservation laws and does not require a penalty term in the final Hamiltonian. Interestingly, in this case, the constraint has a clear physical meaning, namely, that the search space is that of zero $z$-magnetization states~\footnote{The observant reader will notice the existence of another operator that commutes with the total Hamiltonian. This is the parity operator $P=\otimes_{i=1}^n \sigma^x_i$ that splits the Hilbert space to symmetric (eigenvalue $+1$) and antisymmetric (eigenvalue $-1$) states. This operator commutes with the usual transverse-field driver as well.}.  Exploiting conservation laws obviates the need for 
a complete interaction graph, and dramatically reduces the amount of resources needed for the encoding of the graph. Within the standard approach, one would need $\sim n^2$ edges, which on an actual quantum annealer with a bounded degree connectivity would translate to requiring $\sim n^2$ additional qubits~\cite{choi:08,vinci:15}. CQA on the other hand, is considerably less resource demanding, requiring at most $n$ additional edges. This is summarized in Table~\ref{tb:resources}. 

\begin{table}
\begin{tabular}{|c||c|c|}
\hline
  approach & additional & additional qubits \\
     &  edges & after minor embedding \\
\hline
  penalty based & $\sim n^2$ & $\sim n^2$ \\
  CQA & $\sim n$ & $\sim n$ \\
\hline
\end{tabular}
\caption{{\bf Additional resources needed for the encoding of fixed-degree graph partitioning problems.}  While for penalty based approaches the required resources scale quadratically with the original problem size, within CQA the growth is at most linear.}. 
\label{tb:resources}
\end{table}

\begin{figure}[b]
\begin{center}
\includegraphics[width=0.8\columnwidth]{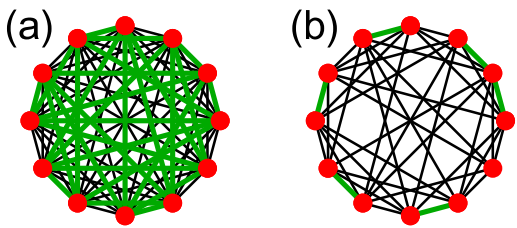}
\caption{(Color online) \textbf{Connectivity of a random $12$-bit degree-6 graph partitioning instance}. The black edges are those of the random instance, whereas the thicker green edges are the additional edges required to satisfy the constraint in the standard approach (left) and in the CQA approach (right).  While `traditional' penalty based embedding requires a fully connected graph, in the CQA approach, only six additional edges are required (so as to close a cycle).}
\label{fig:GP_graphs}
\end{center}
\end{figure}

It is important to note that the above choice of a carefully tailored driver Hamiltonian may potentially come at a cost. One should verify that the minimum gap of the new driver does not scale worse with problem size than the usual transverse-field driver. To test this, we generated many instances of the GP problem and computed the scaling of the typical minimum gap.
We chose random regular graphs of degree $6$ (this subset of problems is known to belong to the NP-complete complexity class). As a first step, we generated hundreds of random instances of different sizes $n$, up to $n=20$. For the scaling of the minimum gap to be meaningful in the context of adiabatic quantum computing, we then handpicked all those instances that had a unique ground state (up to a global bit-flip) so as to make sure that the calculated gap is the ``relevant gap'', i.e., the gap between the ground state and an even excited state in the limit of $s\to 1$ with $s$ being the adiabatic parameter. This was accomplished by exhaustively enumerating all energy levels of the generated instances. 
Figure~\ref{fig:GP_graphs} shows a random $n=12$-bit instance. The figure also shows the required additional edges in the standard approach (left) and in the present CQA approach (right). While in the former method $O(n^2)$ additional edges are needed, in the latter case at most $n$ additional edges are required. 

To calculate the gaps at different points along the annealing path, i.e., different values of $s$ in the parametrized family of Hamiltonians \hbox{$H=(1-s)H_d+s H_p$}, we used exact diagonalization techniques (specifically, the Lanczos algorithm) to calculate the first few energy levels of the total Hamiltonian $H$. In both the standard approach and the CQA, the diagonalization was restricted to the relevant sector in which the evolution takes place. In the former case, this is the symmetric subspace, namely the eigenvalue $1$ subspace of the parity operator $P=\bigotimes_{i=1}^n \sigma^x_i$. For CQA, the relevant subspace is the symmetric subspace within the zero sector of the constraint $C=\sum_i \sigma^z_i$, the space of allowed configurations. An example of the gap calculations for a random $14$-bit instance is given in Fig.~\ref{fig:GPgapn14} below.

\begin{figure}[htp]
\begin{center}
\includegraphics[width=0.85 \columnwidth]{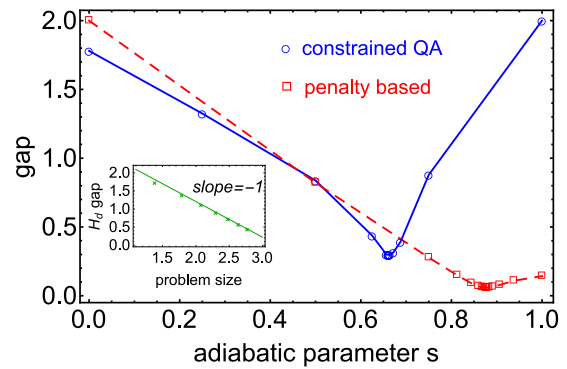}
\caption{(Color online) \textbf{First excitation gap for a random $14$-bit degree-6 graph partitioning instance as a function of the adiabatic parameter $s$.}  The blue curve corresponds to the conservation law-based (CQA) embedding approach whereas the penalty based technique is the dashed red curve. \textbf{Inset: Initial ($H_d$) gap, scaling inversely with problem size.} }
\label{fig:GPgapn14}
\end{center}
\end{figure}

\begin{figure}[htp]
\begin{center}
\includegraphics[width=0.98 \columnwidth]{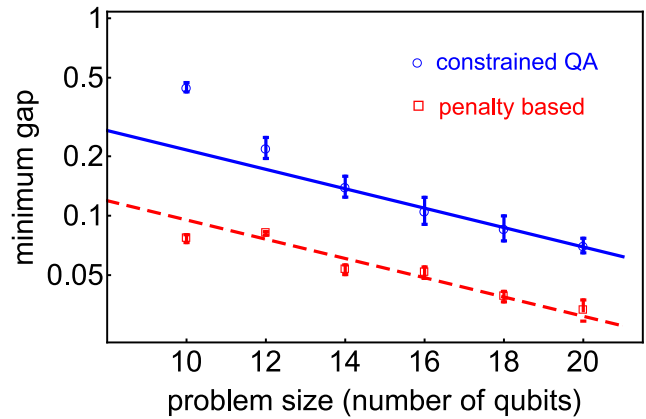}
\caption{(Color online) \textbf{Exponential scaling of the typical minimum relevant gap with problem size for random GP instances (log-linear scale).} The resource-efficient CQA minimum gap (blue circles) scales similarly to that of the penalty based method (red squares), and is typically about three times larger.}
\label{fig:GPgap}
\end{center}
\end{figure}

We then computed the minimum gap as a function of problem size for instances with unique graph partitioning solutions, using traditional embedding and then CQA. In the standard traditional approach, the driver is the usual transverse-field driver, and the problem Hamiltonian is augmented by a penalty term. Since the latter contains $\sim n^2$ terms, the problem Hamiltonian was further normalized by a factor of $1/n$ to make sure that its norm scales linearly with problem size in order to maintain the extensivity of the energy. In our CQA approach, neither penalties nor normalization is required and the cyclic XY model has been used as driver. Figure~\ref{fig:GPgap} shows the results revealing a similar scaling for the two approaches, with the CQA gap being consistently about three times larger.

\subsection{Graph coloring}
\label{GC}

We also considered the problem of graph coloring, that  illustrates the versatility of the proposed technique to address the case of multiple disjoint constraints of the total magnetization type, namely, $\sum_i \sigma^z_i=c$  for a nonzero constant $c$.   
In  graph coloring (GC), one is given an undirected graph $G=(V,E)$, $n = |V|$, and a set of $n_c$ colors, and the question is whether
it is possible to color each vertex in the graph with a specific color, such that no edge connects two vertices of the same color~\cite{karp:72}.
For a GC instance  with $n_c$ colors, we assign $n_c$  spins to each node of $G$, associating $\sigma^z_{i,k}$ to a qubit  
which is `up' if vertex $i$ is colored with color $k$, and is `down' otherwise. The objective function then takes the form~\cite{lucas:14}
\bea
H'_p&=& \sum_{(ij)\in E} \sum_{k=1}^{n_c} (1+\sigma^z_{i,k})(1+\sigma^z_{j,k}) \\\nonumber
&+& \alpha \sum_{i=1}^{n} \left[\sum_{k=1}^{n_c}\sigma^z_{i,k}-(n_c+2)\right]^2 \,.
\eea
The first term assigns a positive cost to every edge that connects two vertices of the same color, while the second term enforces the $n$ constraints that each vertex has exactly one color, and provides an energy penalty each time any of these are violated (here $\alpha$ can be set to $1$).  
A zero-energy ground state implies a solution to the coloring problem on this graph with $n_c$ colors.  
The total number of spins required is thus $n \times n_c$, and the $n_c$ spins for each of the nodes of $G$ form a clique, i.e., a fully connected subgraph. 

As in GP, the cyclic XY driver can be used here as well, except that now there are $n$ independent cycles, each 
associated with a different vertex of the graph, and  composed of the $n_c$ qubits associated with that vertex. In this case, each cycle will impose one of the $n$ constraints.
Such a choice of $H_d$ will then, as before, obviate the need for the $n_c$-sized cliques prescribed by the traditional embedding. 
It is also instructive to consider here
a driver Hamiltonian of the `clique' form, namely, 
 \beq
 H_d=- \frac1{n_c}  \sum_{k=1}^n \sum_{i=0,j<i}^{n_c} \left( \sigma_{k,i}^x \sigma_{k,j}^x + \sigma_{k,i}^y \sigma_{k,j}^y\right) \,,
\eeq
which requires cliques similar to the traditional embedding procedure, but has a constant initial relevant gap and does not require the introduction of  a new energy scale into the problem. Since for each of the $n$ constraints, $c= n_c -2$, the search space here is one in which all the spins belonging to the same vertex but one must be down (imposing one color per vertex). The above driver is effectively a sum of terms of the form $-(M^2_x+M_y^2)$, where $M_{x/y}$ is 
the total magnetization of the spins for each vertex in the $x/y$ direction. The ground state in this case is a direct product of the ground states in the various vertex subsystems, where the latter  is the equal superposition of configurations with only one spin pointing up.

\subsection{Constrained satisfaction problems}
\label{CSP} 

We now illustrate our approach for another important set of problems, namely, that of the constraint satisfaction type, which as we shall show gives rise to a different form of constraints.
Constraint satisfaction problems (CSPs) play a dominant role in quantum annealing optimization both because of their key role in complexity theory as well as their importance in many practical areas. In fact, CSP was the first class of problems to be illustrated for a potential quantum annealing speedup~\cite{farhi:01} and subsequent studies (see Refs.~\cite{young:08,young:10,hen:11,hen:12,farhi:12} and references therein). In a CSP instance there are $n$ bits and $m$ clauses,  where each clause is a logical condition on a small number of randomly chosen bits.  
A configuration of the bits (spins) is a satisfying assignment if it satisfies all the clauses. 
A canonical example of CSP, which we consider here, is that of $3$SAT wherein each clause consists of three bits, and the clause is satisfied for seven of the eight possible 3-bit configurations. 

In the traditional encoding of this type of problem in the quantum annealing framework, each bit variable
is represented in the Hamiltonian by a $\sigma_i^z$ operator, where $i$ labels the spin.  Each clause is then
converted to an energy function which depends on the spins associated with the
clause, such that the energy is zero if the clause is satisfied and is
positive if it is not.  The 
problem Hamiltonian $H_p$ can then be written as a sum of terms, i.e. $H_p = \sum_{m=1}^{M} H^{(m)}$, 
where $m$ is the clause index and $H^{(m)}$ is the energy associated with
the clause. 
The $3$SAT clause Hamiltonian $H^{(m)}$ can be written as
$H^{(m)}=|j_m\rangle \langle j_m|$ where $|j_m\rangle$ is the 3-bit configuration which violates clause $m$. Here, the energy
is zero if the clause is satisfied and is one otherwise. In terms of Pauli operators acting on the bits involved in the clause, the Hamiltonian can be written using three-spin interactions as:
\beq
H^{(m)}= \frac1{8}\bigotimes_{i=1}^{3} \left[1+(-1)^{b_{i,m}} \sigma^z_i\right] \,,
\eeq
where $b_{i,m}$ are the values of the bits comprising $|j_m\rangle$ and $i=1,2,3$ labels the spins in the clause.

Problems of the $3$SAT type can be viewed as unconstrained, with a Hamiltonian that is simply a sum of penalties $H^{(m)}$. To apply the techniques presented here
we shall treat some of these penalties as constraints that will then turn into conserved 
quantities (let us denote this subset of clauses by $C$) while the others remain a part of the Hamiltonian. We shall require that 
clauses identified as constraints involve mutually disjoint sets of spins. Within the new approach, the problem Hamiltonian will consist 
only of `non-constraint' clauses, explicitly  $H_p=\sum_{m \notin C} H^{(m)}$ and as many of them as possible. This can be done due 
to the other constraints becoming immediately 
conserved operators provided a suitable driver is found.
As for the driver Hamiltonian, we will choose for each clause $m \in C$ the driver
\bea
H_d^{(m)}&=&-\sum_{i \neq j_m} |i\rangle \left( \sum_{i' \neq i,j_m} \langle i'| \right) =1\\\nonumber
&-& 8|+\rangle\langle+|+2\sqrt{2}\left( |j_m\rangle\langle +| + |+\rangle \langle j_m|\right)-2 |j_m\rangle\langle j_m|\,,
\eea
where $H_d^{(m)}$ acts on the three bits in the clause and $|+\rangle$ is the fully-symmetric state. This driver, similarly to the problem Hamiltonian, involves $3$-spin operators and can be written in terms of Pauli matrices since  
\bea
|+\rangle\langle +|&=& \frac1{8}\bigotimes_{i=1}^{3} \left[1+\sigma^x_i\right] \quad \textrm{and}\\\nonumber 
|j_m\rangle\langle +| &=& \frac1{8\sqrt{8}}\bigotimes_{i=1}^{3} \left[1+\sigma^x_i +(-1)^{b_{i,m}} (\rmi \sigma^y_i +\sigma^z_i)\right]\,.
\eea
The above driver ensures that the evolution is restricted to the subspace of configurations simultaneously satisfying  {\it all} of the constraint clauses. 
For all bits that are not present in the constraint clauses (let us label this set of bits $K$), a transverse-field term may be chosen. Therefore, the total driver 
Hamiltonian will be
\beq
H_d=\sum_{m \in M} H_d^{(m)} - \sum_{k \in K} \sigma^x_k\\,
\eeq
with the corresponding ground state
\hbox{$
|\psi\rangle = \bigotimes_{m \in C} (\frac1{\sqrt{7}} \sum_{i \neq j_m} | i \rangle) \bigotimes_{k \in K} |+\rangle_k
$}.

It is worth noting that other problems not discussed here may also benefit from suitably chosen driver Hamiltonians. One such example has been studied in Ref.~\cite{martonak:04} in the context of the traveling salesman problem, where four-body terms were used to construct the driver Hamiltonian.

\section{Conclusions}

The method described in this study addresses one of the major obstacles to solving combinatorial optimization problems using quantum annealers and in a sense partially resolves it. We have shown that by choosing a suitable driver, one can exploit symmetries of the quantum annealing Hamiltonian to
obviate the need for penalty terms to impose constraints as these are naturally enforced by the dynamics in the new approach. The removal of penalty terms in turn eliminates the required increased connectivity and the shrinking of the dynamical range of the interactions. 
Interestingly, the current approach is inherently quantum and does not have a classical counterpart. As such, it may be viewed as providing a form of a `purely quantum enhancement', which could become important in light of recent results pointing towards the significance of tunneling in QA processes~\cite{googleTunnelingII}.

As we have demonstrated, a smart choice of initial Hamiltonian translates, in general, into a considerabe reduction in resources, at times in orders of magnitude as in the case of graph partitioning. The significance of the method proposed here is therefore not only of theoretical interest but of potentially immense practical significance. It touches deeply on the scales of problems that can actually be embedded to benchmark and evaluate experimental quantum annealers leading to capabilities far beyond currently testable problem sizes. As such the present method is expected to substantially contribute to the benchmarking capabilities of near-future experimental quantum annealers potentially leading to exciting new discoveries of quantum annealing speedups. 

It remains to be seen how easy it is to experimentally engineer interactions of the form discussed above or any other useful interactions that naturally impose constraints.  
Even though the usual transverse-field driver Hamiltonian may be easier to implement,  these slightly more complex interaction terms can help to avoid  some of the obstacles faced by practical quantum annealers. It is therefore not unreasonable to believe  that the implementation of driver Hamiltonians of the forms suggested above can have a considerable impact on the engineering and, in turn, the success of experimental quantum annealers in dealing with encoding and solution of combinatorial optimization problems.

Another problem, which so far has  been suffering from the inability of quantum annealers to successfully deal with external constraints, is that of graph minor embedding~\cite{choi:08,vinci:15}. There, one is asked to embed a given target graph on a usually larger, more sparse hardware graph by assigning several vertices of the latter to each vertex of the former. Imposing the same orientation on all physical bits corresponding to the same logical bit is a bonafide constraint which is normally dealt with using penalty terms. An interesting direction of research would be to look for suitable driver Hamiltonians that can resolve the constraint-related handicaps of graph minor embedding as well. 

The above approach also brings to mind other physical systems that may be utilized towards solving other constrained optimization problems via quantum annealing. 
One such system is that of ultra-cold bosons placed in optical lattices. There, one can smoothly `anneal' the system from a delocalized superfluid to a Mott insulator through a quantum phase transition and vice versa. In the course of the evolution, the number of particles is obviously a conserved quantity. An intriguing concept would be to utilize systems undergoing superfluid to Mott insulator phase transitions for computation, i.e., for solving appropriate constrained optimization problems. 

Another aspect that is important to study is the performance of the above approach in non-ideal settings, i.e., at finite temperatures and in the presence of the decohering effects of the environment and other sources of noise and inaccuracies. 
As this work is meant to serve as a `proof of concept,' these aspects will be studied elsewhere. 

\section{Acknowledgments}

We thank Mohammad Amin, Sergio Boixo and Daniel Lidar for useful comments and discussions. IH acknowledges support by ARO grant number W911NF-12-1-0523. FMS acknowledges support from NSF, under grant CCF-1551064. Computation for the work described here was supported by the University of Southern California's Center for High-Performance Computing (\url{http://hpcc.usc.edu}).

\bibliography{refs}
\end{document}